\documentclass[twocolumn,nofootinbib,amsmath,amssymb,superscriptaddress,showpacs,pre]{revtex4-1}
\usepackage{graphicx}
\usepackage{epstopdf}
\usepackage{bm}
\usepackage{latexsym}

\usepackage{rotating}

\newcommand{\smin}{s_{min}}

\newcommand{\w}{\omega}

\begin{document}
\title{An empirical study of the tails of mutual fund size}
\author{Yonathan \surname{Schwarzkopf}}
\affiliation{California Institute of Technology, Pasadena, CA 91125.}
\affiliation{ Santa Fe Institute, Santa Fe, NM 87501.} 
\author{J. Doyne \surname{Farmer}$^{2,}$}
\affiliation{Luiss Guido Carli, ROMA Italy.}%

\date{\today}

\begin{abstract}
The mutual fund industry manages about a quarter of the assets in the U.S. stock market and thus plays an important role in the U.S. economy.  The question of how much control is concentrated in the hands of the largest players is best quantitatively discussed in terms of the tail behavior of the mutual fund size distribution.  We study the distribution empirically and show that the tail is much better described by a log-normal than a power law, indicating less concentration than, for example, personal income.  The results are highly statistically significant and are consistent across fifteen years.  This contradicts a recent theory concerning the origin of the power law tails of the trading volume distribution.  Based on the analysis in a companion paper, the log-normality is to be expected, and indicates that the distribution of mutual funds remains perpetually out of equilibrium.
\end{abstract}


\maketitle

\section{Introduction}\label{section_introduction}

As of 2007 the mutual fund industry controlled $23\%$ of household taxable assets in the United States\footnote{
Data is taken from the Investment Company Institute's 2007 fact book available at www.ici.org. }. 
In absolute terms this corresponded to 4.4 trillion USD and 24\% of U.S. corporate equity holdings.   Large players such as institutional investors are known to play an important role in the market \citep{corsetti-2001}.  This raises the question of who has this influence:  Are mutual fund investments concentrated in a few dominant large  funds, or spread across many funds of similar size?  Are there mutual funds that are so large that they are ``too big to fail"?

This question is best addressed in terms of the behavior of the upper tail of the mutual fund size distribution. The two competing hypotheses usually made in studies of firms are Zipf's law vs. a lognormal.  Zipf's law means that the distribution of the size $s$ is a power law with tail exponent $\zeta_s \approx 1$, i.e. 
\[
P(s>X) \sim X^{-\zeta_s},
\]
Log-normality means that $\log s$ has a normal distribution, i.e. the density function $p_{LN}(s)$ obeys 
\[
p(s)=\frac{1}{s\sigma\sqrt{2\pi}}\exp\left(-\frac{(\log(s)-\mu_s)^2}{2\sigma_s^2}\right).
\]

From the point of view of extreme value theory this distinction is critical, since it implies a completely different class of tail behavior\footnote{
According to extreme value theory a probability distribution can have only four possible types of tail behavior.  The first three correspond to distributions with finite support, thin tails, and tails that are sufficiently heavy that some of the moments do not exist, i.e. power laws. The fourth category corresponds to distributions that in a certain sense do not converge; it is remarkable that most known distributions fall into one of the first three categories \citep{Embrechts97}.}.
These are both heavy tailed, but Zipf's law is much more heavy tailed.  For a log-normal all the moments exist, whereas for Zipf's law none of the moments exist.  For Zipf's law an estimator of the mean fails to converge.  In practical terms, for mutual funds this would imply that for any sample size $N$, with significant  probability an individual fund can be so large that it is bigger than all other $N-1$ firms combined.  In contrast, for a log-normal, in the limit as $N \to \infty$ the relative size of a single fund becomes negligible.

This question takes on added meaning because the assumption that mutual funds follow Zipf's law has been argued to be responsible for the observed power law distribution of trading volume \citep{levy-1996,solomon-2001}.  Gabaix et al. have also asserted that the mutual fund distribution follows Zipf's law and have used this in a proposed explanation for the distribution of price returns \citep{gabaix-2003-nature,gabaix-2006}.

We resolve this empirically using the Center for Research in Security Prices (CRSP) dataset and find that the equity fund size distribution is much better described by a log-normal distribution. 

Our results are interesting in the broader context of the literature on firm size.  Mutual funds provide a particularly good type of firm to study because there are a large number of funds and their size is accurately recorded.   It is generally believed that the resulting size distribution from aggregating across industries has a power law tail that roughly follows Zipf's law, but for individual industries the tail behavior is debated\footnote{
Some studies have found that the upper tail is a log-normal \citep{simon-1958,stanley-1995,ijiry-1977,stanley-1996,amaral-1997,bottazzi-2003a,dosi-2005}  while others have found a power law \citep{axtell-2001,bottazzi-2003a,dosi-2005}}.
 A large number of stochastic process models have been proposed to explain this\footnote{
 For past stochastic models  see \citep{gibrat-1931,simon-1955,simon-1958,mandelbrot-1963,ijiry-1977,sutton-1997,gabaix-2003-mit,gabaix-2003-nature}}.
  Our results add support to the notion that for single industries the distribution is log-normal.
  
 The log-normality of the distribution of mutual funds is also interesting for what it suggests about the underlying processes that determine mutual fund size.  In a companion paper \cite{Schwarzkopf10b} we develop a model for the random process of mutual fund entry, exit and growth under the assumption of market efficiency, and show that this gives a good fit to the data studied here.  We show that while the steady-state solution is a power law, the timescale for reaching this solution is very slow.  Thus given any substantial non-stationarity in the entry and exit processes the distribution will remain in its non-equilibrium log-normal state.  See the discussion in Section V.

\section{Data Set}\label{data_set}

We analyze the Center for Research in Security Prices (CRSP) Survivor-Bias-Free US Mutual Fund Database\footnote{
The US Mutual Fund Database can be purchased from the Center for Research in Security Prices (www.crsp.com).}.
The database is survivor bias free as it contains historical performance data for both active and inactive mutual funds.
We study monthly data from 1991 to 2005\footnote{
There is data on mutual funds starting in 1961, but prior to 1991 there are very few entries.  There is a sharp increase in 1991, suggesting incomplete data collection prior to 1991.}
on all reported equity funds.
We define an equity fund as one whose portfolio consists of at least $80\%$ stocks.  The 
results are not qualitatively sensitive to this, e.g. we get essentially the same results even if we use all funds.    
The data set has monthly values for the Total Assets Managed (TASM) by the fund and the Net Asset Value (NAV).  We define the size $s$ of a fund to be the value of the TASM, measured in millions of US dollars and corrected for inflation relative to July 2007.  Inflation adjustments are based on the Consumer Price Index, published by the BLS.

\section{Is the tail a power law?}\label{section_is_pl}
\begin{figure}
\begin{center}
\includegraphics[width=9cm]{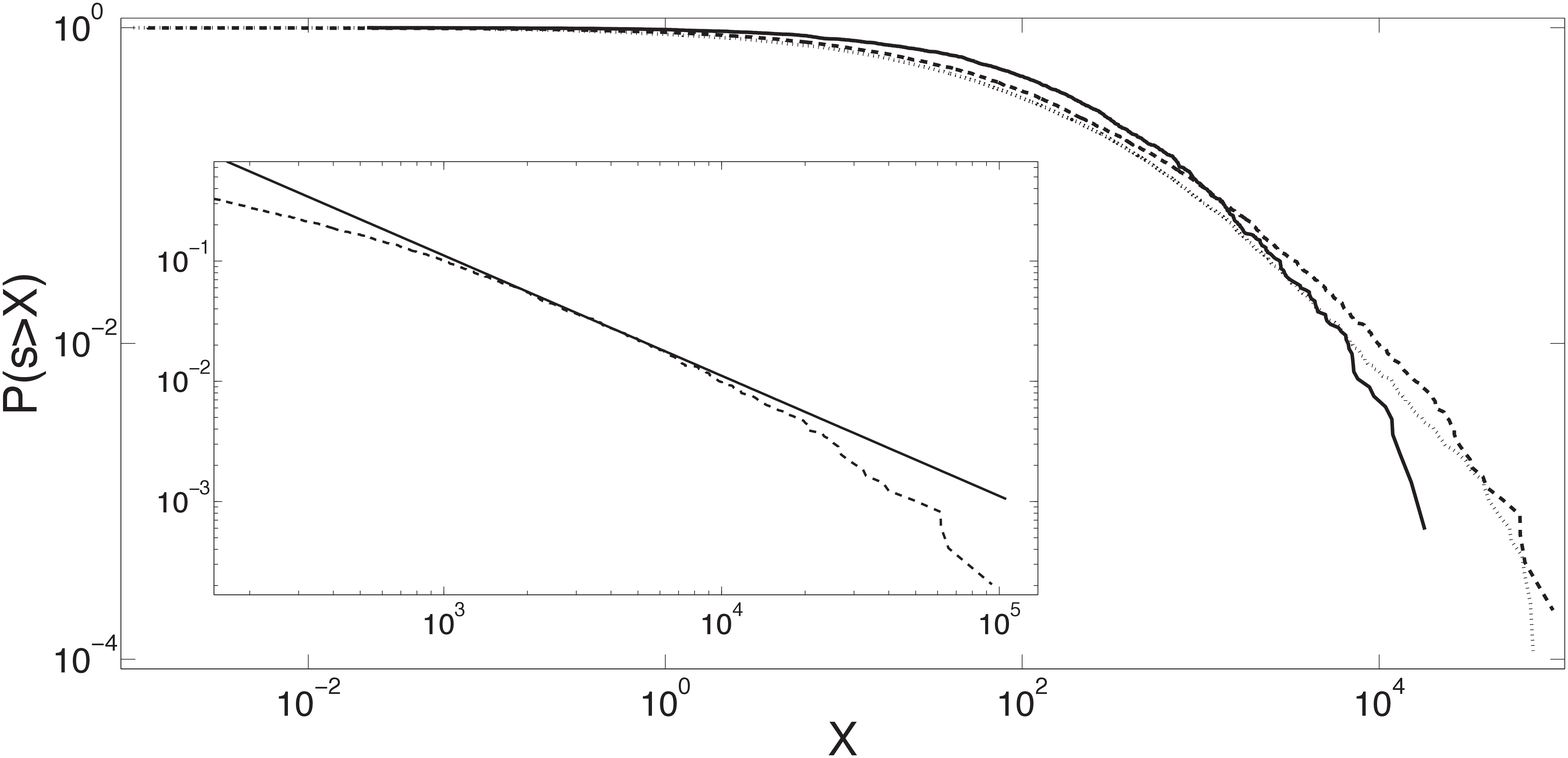}
\caption{\label{s_CDF}
The CDF for the mutual fund size $s$ (in millions of 2007 dollars) is plotted with a double logarithmic scale. The cumulative distribution for funds existing at the end of the years 1993, 1998 and 2005 are given by the full, dashed and dotted lines respectively.\newline
 Inset: The upper tail of the CDF for the mutual funds existing at the end of 1998 (dotted line)  is compared to an algebraic relation with exponent $-1$  (solid line).}
\end{center}
\end{figure}
Despite the fact that the mutual fund industry offers a large quantity of well-recorded data, the size distribution of mutual funds has not been rigorously studied. This is in contrast with other types of firms where the size distribution has long been an active research subject. The fact that the distribution is highly skewed and heavy tailed can be seen in Figure~\ref{s_CDF}, where we plot the cumulative distribution of sizes $P(s>X)$ of mutual fund sizes in three different years.

A visual inspection of the mutual fund size distribution suggests that it does not follow Zipf's law\footnote{
 Previous work on the size distribution of mutual funds by Gabaix et al. \citep{gabaix-2003-mit,gabaix-2003-nature,gabaix-2006}
  argued for a power law while we argue here for a log-normal.}.
In the inset of Figure~\ref{s_CDF} we compare the tail for funds with sizes $s>10^2$ million to a power law $s^{-\zeta_s}$, with $\zeta_s=-1$.  Whereas a power law corresponds to a straight line when plotted on double logarithmic scale, the data show substantial and consistent downward curvature. The main point of this paper is to make more rigorous tests of the power law vs. the log-normal hypothesis.   These back up the intuitive impression given by this plot, indicating that the data are not well described by a power law. 

To test the validity of the power law hypothesis we use the method developed by Clauset et al. \cite{clauset-2007}. 
They use the somewhat strict definition\footnote{
In extreme value theory a power law is defined as any function that in the limit $s \to \infty$ can be written $p(s) = g(s)s^{-(\zeta_s + 1)}$ where $g(s)$ is a slowly varying function.  This means it satisfies $\lim_{s \to \infty} g(ts)/g(s) =  C$ for any $t > 0$, where $C$ is a positive constant.  The test for power laws in reference \citep{clauset-2007} is too strong in the sense that it assumes that there exists an $s_0$ such that for $s > s_0$, $g(s)$ is constant.}
that the probability density function $p(s)$ is a power law if there exists an $\smin$ such that for sizes larger than $\smin$, the functional form of the density $p(s)$ can be written
\begin{equation}\label{eq_pdf_pl}
p(s)=\frac{\zeta_s}{\smin}\left(\frac{s}{\smin}\right)^{-(\zeta_s + 1)},
\end{equation}
where the distribution is normalized in the interval $[\smin,\infty)$. 
There are two free parameters $\smin$ and $\zeta_s$.  
This crossover size $\smin$ is chosen such that it minimizes 
the Kolmogorov-Smirnov (KS) statistic $D$, which is the distance between  the 
CDF of the empirical data  $P_{e}(s)$ and that of the fitted model $P_f(s)$, i.e.
\[
D=\max_{s\geq \smin}\left| P_e(s)-P_f(s)\right|.
\]

Using this procedure we estimate $\zeta_s$ and $\smin$ for the years 1991- 2005 as shown in Table~\ref{table}.  The values of $\zeta_s$ computed in each year range from $0.78$ to $1.36$ and average $\bar{\zeta_s}= 1.09\pm 0.04$.  If indeed these are power laws this is consistent with Zipf's law.  But of course, merely computing an exponent and getting a low value does not mean that the distribution is actually a power law.


To test the power law hypothesis more rigorously we follow the Monte Carlo method utilized by Clauset et al.   Assuming independence, for each year we generate $10,000$ synthetic data sets,
each drawn from a power law with the empirically measured values of $\smin$ and $\zeta_s$.  For each data-set we calculate the KS statistic to its best fit. The $p$-value is the fraction of the data sets for which the KS statistic to its own best fit is larger than the KS statistic for the empirical data and its best fit.


The results 
are summarized in Table~\ref{table}.  The power law hypothesis is rejected with two standard deviations or more in six of the years and rejected at one standard deviation or more in twelve of the years (there are fifteen in total).  Furthermore there is a general pattern that as time progresses the rejection of the hypothesis becomes stronger.  We suspect that this is because of the increase in the number of equity funds.  As can be seen in Table~\ref{table}, 
 the total number of equity funds increases roughly linearly in time, and the number in the upper tail $N_{tail}$ also increases.
 
\begin{turnpage}
\begin{table*}
\begin{center}
\footnotesize
\begin{tabular}{|c|c|c|c|c|c|c|c|c|c|c|c|c|c|c|c||c|c|}
\hline
variable & 91&92&93&94&95&96&97&98&99&00&01&02&03&04&05&mean&std \\
\hline \hline
${\cal R}$ &  -0.50 &  -1.35 &  -1.49& -1.71 & -3.29 & -18.42 &  -2.25 &  -1.29 &  -6.57 &  -4.96&   -2.63  & -2.95 &  -2.00 &  -1.05& -0.99 & -3.43 & 4.45  \\
\hline \hline
$N$ & 372 & 1069 & 1509 & 2194 & 2699  &  3300 & 4253 & 4885 & 5363 & 5914 & 6607 & 7102
&7794& 8457 & 8845& - & -  \\
\hline \hline
E$[s]$ (mn)&810&  385  &  480   & 398   & 448  &  527  &  559  &  619  &  748  &  635  &  481  &  335  & 425  &  458 &  474 & 519 & 134\\
 \hline 
 Std$[s]$ (bn) &1.98   & 0.99  & 1.7  &  1.66 & 1.68  &  2.41  &  2.82  &  3.38   & 4.05  &  3.37 &  2.69  &  1.87  & 2.45  &  2.64   & 2.65 & 2.42& 0.8\\
\hline \hline
E$[\w]$&5.58  &  4.40  &  4.40   & 3.86   & 3.86  &  3.91  &  3.84  &  3.85  &  4.06  &  3.97  &  3.60  &  3.37  & 3.55  &  3.51  &  3.59&3.96&0.54\\
 \hline 
 Std$[\w]$ &1.51   & 1.98  &  2.09  &  2.43 & 2.50  &  2.46  &  2.50  &  2.51   & 2.46  &  2.45 &  2.63  &  2.42  & 2.49  &  2.59   & 2.50 & 2.34& 0.29\\
\hline \hline
$\zeta_s$ & 1.33 & 1.36 & 1.19 & 1.15 & 1.11 & 0.78 & 1.08 &1.10 & 0.95 & 0.97 & 1.01 & 1.07 & 1.07 &  1.10 & 1.14 & 1.09 & 0.14 \\
\hline
$\smin$ & 955 & 800 & 695 & 708 & 877 & 182 & 1494 & 1945 &  1147 & 903 & 728 & 836 & 868 & 1085 & 1383 & 974 & 408  \\
\hline $N_{tail}$& 81 & 129 & 232 & 256 & 280 & 1067&  290 &  283 & 557 & 662 & 717& 494 & 652 & 630 & 550 & - & -  \\
\hline \hline
$p$-value & 0.58 &   0.48 &    0.02 &   0.45 &    0.07 & 0 & 0.01 &  0.11 &  $5 \,10^{-4}$ & 0.04 & 0.03 &  0.07 &  0.08 &    0.08 & 0.15 & 0.15 & 0.19  \\
\hline \hline
\end{tabular}
\normalsize 
\end{center}
\label{default}
\caption{\label{table}
Table of monthly parameter values for equity funds defined such that the portfolio contains a fraction of at least $80\%$ stocks. 
The values for each of the monthly parameters (rows) were calculated for each year (columns). The mean and standard deviation are evaluated for the monthly values in each year.  \newline
${\cal R}$ - the  base 10 log likelihood ratio of a power law fit relative to a log-normal fit as given by equation (\ref{eq_LLR}).
 A negative value of ${\cal R}$ indicates that the log-normal hypothesis is a likelier description than a power law. For all years the value is negative meaning that the log-normal distribution is more likely.  \newline
$N$ - the number of equity funds existing at the end of each year. \newline
$E[\w]$ - the mean log size of funds existing at the end of each year. \newline
$Std[\w]$ - the standard deviation of log sizes for funds existing at the end of each year. \newline
$E[s]$ - the mean size (in millions) of funds existing at the end of each year. \newline
$Std[s]$ - the standard deviation of sizes (in billions) for funds existing at the end of each year. \newline
$\zeta_s$ - the power law tail exponent (\ref{eq_pdf_pl}). \newline
$\smin$ - the lower tail cutoff (in millions of dollars) above which we fit a power law  (\ref{eq_pdf_pl}). \newline
$N_{tail}$ - the number of equity funds belonging to the upper tail s.t. $s\geq \smin$. \newline
$p$-value - the probability of obtaining a goodness of fit at least as bad as the one calculated for the empirical data, under the null hypothesis of a power law upper tail. \newline
 }
\end{table*} 
\end{turnpage}

We conclude that the power law tail hypothesis is questionable but cannot be unequivocally rejected in every year.  Stronger evidence against it comes from comparison to a log-normal, as done in the next section.

\section{Is the tail log-normal?}\label{section_is_ln}
\begin{figure}
\begin{center}
\includegraphics[width=8.5cm]{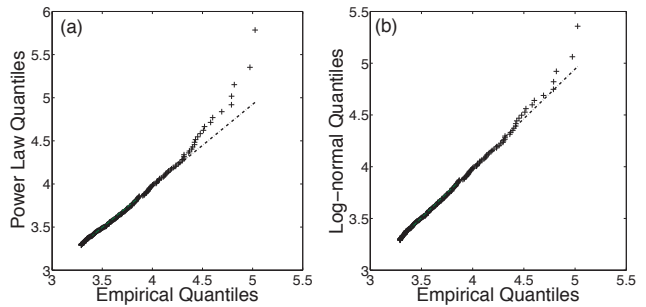}
\caption{\label{qq-pl}
A Quantile-Quantile (QQ)  plot for the upper tail of the size distribution of equity funds. The quantiles are the base ten logarithm of the fund size, in millions of dollars.  The empirical quantiles are calculated from the size distribution of funds existing at the end of the year 1998. The empirical data were truncated from below such that only funds with size $s\geq \smin$ were included in the calculation of the quantiles.   (a) A QQ-plot with the empirical quantiles as the x-axis and the quantiles for the best fit  power law as the y-axis. The power law fit for the data was done using the maximum likelihood described in Section~\ref{section_is_pl}, yielding $\smin=1945$ and $\alpha= 1.107$.  
(b) A QQ-plot with the empirical quantiles as the x-axis and the quantiles for the best fit  log-normal as the y-axis, with the same $s_{min}$ as in (a).  
 The log-normal fit for the data was done used the maximum likelihood estimation given $\smin$ (\ref{eq_pln_smin})
 yielding $\mu=2.34$ and $\sigma=2.5$.}
\end{center}
\end{figure} 
A visual comparison between the two hypotheses can be made by looking at the Quantile Quantile (QQ) plots for the empirical data compared to each of the two hypotheses. In a  QQ-plot we plot the quantiles of one distribution as the x-axis and the other's as the y-axis. If the two distributions are the same then we expect the points to fall on a straight line.   Figure~\ref{qq-pl} compares the two hypotheses, making it clear that the log-normal is a much better fit than the power law.
For the log-normal QQ plot most of the large values in the distribution fall on the dashed line corresponding to a log-normal distribution, though  the very largest values are somewhat above the dashed line.   This says that the empirical distribution decays slightly faster than a log-normal.  There are two possible interpretations of this result:  Either this is a statistical fluctuation or the true distribution really has slightly thinner tails than a log-normal.  In any case, since a log-normal decays faster than a power law, it strongly suggests that the power law hypothesis is incorrect and the log-normal distribution is a better approximation.

A more quantitative method to address the question of which hypothesis better describes the data is to compare the likelihood of the observation in both hypotheses \citep{clauset-2007}. 
We define the likelihood for the tail of the distribution to be 
\[
L=\prod_{s_j\geq \smin}p(s_j).
\]
We define the power law likelihood as \mbox{$L_{PL}=\prod_{s_j\geq \smin}p_{PL}(s_j)$}
with the probability density of the power law tail given by (\ref{eq_pdf_pl}). 
The lognormal likelihood  is defined as $L_{LN}=\prod_{s_j\geq \smin}p_{LN}(s_j)$
with the probability density of the lognormal  tail given by 
\begin{eqnarray}\label{eq_pln_smin}
p_{LN}(s)&=&\frac{p(s)}{1-P(\smin)} \\%
&=&\frac{\sqrt2}{s\sqrt{\pi}\sigma}\left[\mathrm{erfc}\left(  \frac{\ln \smin -\mu}{\sqrt{2}\sigma} \right)\right]^{-1} \exp\left[-\frac{(\ln s-\mu)^2}{2\sigma^2}\right]. \nonumber
\end{eqnarray}

\begin{figure}
\begin{center}
\includegraphics[width=8.5cm]{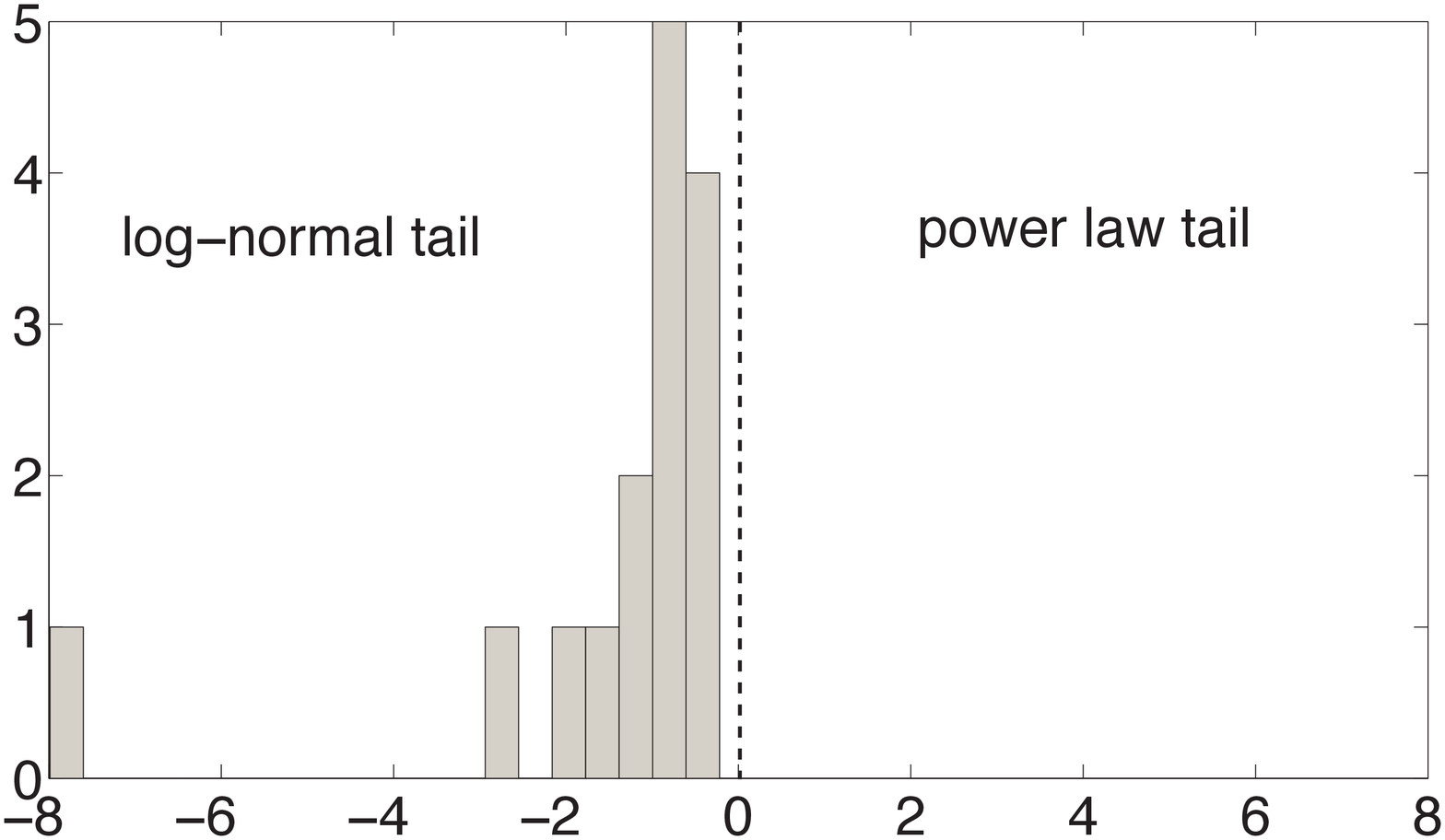}
\caption{\label{logR_hist}
A histogram of the base 10 log likelihood ratios ${\cal R}$ computed using  (\ref{eq_LLR})
for each of the years 1991 to 2005. A negative log likelihood ratio implies that it is more likely that the empirical distribution is log-normal then a power law.  The log likelihood ratio is negative in every year, in several cases strongly so.}
\end{center}
\end{figure}
The more probable that the empirical sample is drawn from a given distribution, the larger the likelihood for that set of observations. The ratio indicates which distribution the data are more likely drawn from.  
We define the log likelihood ratio as 
\begin{equation}\label{eq_LLR}
{\cal R}=\ln\left(\frac{L_{PL}}{L_{LN}}\right).
\end{equation}    
For each of the years 1991 to 2005  we computed the maximum likelihood estimators for both the power law fit and the log-normal fit to the tail, as explained above and in Section~\ref{section_is_pl}. Using the fit parameters, the log likelihood ratio was computed and the results are summarized graphically in Figure~\ref{logR_hist} and in Table~\ref{table}. The ratio is always negative, indicating that the likelihood for the log-normal hypothesis is greater than that of the power law hypothesis in every year.  It seems clear that tails of the mutual fund data are much better described by a log-normal than by a power law.

\section{Implications of log-normality}

The log-normal nature of the size distribution has important implications on the role investor behavior plays in the mutual fund industry. Is the size distribution of mutual funds, i.e. the concentration of assets, determined through investor choice or is it just a consequence of the random nature of the market?  In a companion paper \cite{Schwarzkopf10b} we propose that the size distribution can be explained by a simple random process model. This model, characterizing the entry, exit and growth of mutual funds as a random process, is based on market efficiency, which dictates that fund performance is size independent and fund growth is essentially random. 
This model provides a good explanation of the concentration of assets, suggesting that other effects, such as transaction costs or the behavioral aspects of investor choice, play a smaller role.
  
The fact that the fund distribution is a log-normal is interesting because, as we argue in the companion paper, this indicates a very slow convergence toward equilibrium.   There we find a  time-dependent solution for the underlying random process of mutual fund entry, exit, and growth, and show that the size distribution evolves from a log-normal  towards a Zipf power law distribution.  However, the relaxation to the steady-state solution is extremely slow, with time scales on the order of a century or more.  Given that the mutual fund industry is still young, the distribution remains in its non-equilibrium state as a log-normal.  Furthermore, given that the properties of the entry and exit processes are not stable over long periods of time, the non-equilibrium log-normal state will very likely persist indefinitely.

\section{Conclusions}

We have shown in unequivocal terms that the mutual fund size distribution is much closer to a log-normal than to a power law.  Thus, while the distribution is concentrated, it is not nearly as concentrated as it might be.  Among other things this suggests that that the power law distribution observed for trading volume by Gopikrishnan et al. \cite{Gopikrishnan00} cannot be explained based on a power law distribution for funds.  The companion paper discussed in the previous section \cite{Schwarzkopf10b} constructs a theory that explains the log-normality based on the random nature of the mutual fund entry, exit and growth, and the very long-time scales required for convergence to the steady-state power law solution.

\begin{acknowledgments}
We would like to thank A. Clauset and C. R. Shalizi for useful comments. YS would like to thank Mark B. Wise. We gratefully acknowledge financial support from NSF grant HSD-0624351.  Any opinions, findings and conclusions or recommendations expressed in this material are those of the authors and do not necessarily reflect the views of the National Science Foundation.
\end{acknowledgments}

%

\end{document}